\documentclass[twocolumn]{aastex631}

\newcommand{\source}{PSR~J0311$+$1402}

\newcommand{\gleamx}{GLEAM-X~J162759.5$-$523504.3}
\newcommand{\gpm}{GPM~J1830$-$10}
\newcommand{\meerkatulp}{PSR~J0901$-$4046}

\newcommand{\askapulp}{ASKAP~J1935$+$2148}

\newcommand{\andyulp}{ASKAP~J1832$-$0911}
\newcommand{\lofarulp}{PSR~J0250$+$5854}

\newcommand{\arsco}{AR~Sco}
\newcommand{\pelisoli}{J1912$-$4410}

\usepackage{amsmath}
\usepackage{CJKutf8}
\newcommand{\chinese}[1]{\begin{CJK*}{UTF8}{gbsn}#1\end{CJK*}}

\newcommand{\code}[1]{\textsc{#1}}
\newcommand{\add}[1]{{#1}}

\begin{document}

\title{The discovery of a 41\,s radio pulsar \source{} with ASKAP}

\correspondingauthor{Yuanming Wang}
\email{yuanmingwang@swin.edu.au}

\author[0000-0003-0203-1196]{Yuanming Wang (\chinese{王远明})}
\affiliation{Centre for Astrophysics and Supercomputing, Swinburne University of Technology, Hawthorn, VIC 3122, Australia}
\affiliation{ARC Centre of Excellence for Gravitational Wave Discovery (OzGrav), Hawthorn, VIC 3122, Australia}

\author[0000-0002-2346-6853]{Pavan~A.~Uttarkar}
\affiliation{Centre for Astrophysics and Supercomputing, Swinburne University of Technology, Hawthorn, VIC 3122, Australia}

\author[0000-0002-7285-6348]{Ryan M. Shannon}
\affiliation{Centre for Astrophysics and Supercomputing, Swinburne University of Technology, Hawthorn, VIC 3122, Australia}
\affiliation{ARC Centre of Excellence for Gravitational Wave Discovery (OzGrav), Hawthorn, VIC 3122, Australia}

\author[0009-0001-1631-7462]{Yu Wing Joshua Lee}
\affiliation{Sydney Institute for Astronomy, School of Physics, University of Sydney, NSW 2006, Australia}
\affiliation{Australia Telescope National Facility, CSIRO, Space and Astronomy, PO Box 76, Epping, NSW 1710, Australia}
\affiliation{ARC Centre of Excellence for Gravitational Wave Discovery (OzGrav), Hawthorn, VIC 3122, Australia}

\author[0000-0003-0699-7019]{Dougal Dobie}
\affiliation{Sydney Institute for Astronomy, School of Physics, University of Sydney, NSW 2006, Australia}
\affiliation{ARC Centre of Excellence for Gravitational Wave Discovery (OzGrav), Hawthorn, VIC 3122, Australia}

\author[0000-0002-2066-9823]{Ziteng Wang}
\affiliation{International Centre for Radio Astronomy Research, Curtin University, Bentley, WA 6102, Australia}

\author[0000-0003-2149-0363]{Keith W. Bannister}
\affiliation{Australia Telescope National Facility, CSIRO, Space and Astronomy, PO Box 76, Epping, NSW 1710, Australia}
\affiliation{Sydney Institute for Astronomy, School of Physics, University of Sydney, NSW 2006, Australia}

\author[0000-0002-4079-4648]{Manisha Caleb}
\affiliation{Sydney Institute for Astronomy, School of Physics, University of Sydney, NSW 2006, Australia}

\author[0000-0001-9434-3837]{Adam T.~Deller}
\affiliation{Centre for Astrophysics and Supercomputing, Swinburne University of Technology, Hawthorn, VIC 3122, Australia}
\affiliation{ARC Centre of Excellence for Gravitational Wave Discovery (OzGrav), Hawthorn, VIC 3122, Australia}

\author[0000-0002-5067-8894]{Marcin Glowacki}
\affiliation{Institute for Astronomy, University of Edinburgh, Royal Observatory, Edinburgh, EH9 3HJ, United Kingdom}
\affiliation{International Centre for Radio Astronomy Research, Curtin University, Bentley, WA 6102, Australia}
\affiliation{Inter-University Institute for Data Intensive Astronomy, Department of Astronomy, University of Cape Town, Cape Town, South Africa}

\author[0000-0003-4193-6158]{Joscha N. Jahns-Schindler}
\affiliation{Centre for Astrophysics and Supercomputing, Swinburne University of Technology, Hawthorn, VIC 3122, Australia}
\affiliation{ARC Centre of Excellence for Gravitational Wave Discovery (OzGrav), Hawthorn, VIC 3122, Australia}

\author[0000-0002-2686-438X]{Tara Murphy}
\affiliation{Sydney Institute for Astronomy, School of Physics, University of Sydney, NSW 2006, Australia}
\affiliation{ARC Centre of Excellence for Gravitational Wave Discovery (OzGrav), Hawthorn, VIC 3122, Australia}

\author[0000-0001-8057-0633]{Reshma Anna-Thomas}
\affiliation{ASTRON, Netherlands Institute for Radio Astronomy, Oude Hoogeveensedĳk 4, 7991 PD Dwingeloo, The Netherlands}
\affiliation{Anton Pannekoek Institute for Astronomy, University of Amsterdam, Science Park 904, 1098 XH Amsterdam, The Netherlands}

\author[0000-0002-8383-5059]{N.~D.~R.~Bhat}
\affiliation{International Centre for Radio Astronomy Research, Curtin University, Bentley, WA 6102, Australia}

\author{Xinping Deng}
\affiliation{Hebei Key Laboratory of Radio Astronomy Technology, Shijiazhuang, 050081, Hebei, People's Republic of China}
\affiliation{Australia Telescope National Facility, CSIRO, Space and Astronomy, PO Box 76, Epping, NSW 1710, Australia}

\author[0000-0001-9817-4938]{Vivek Gupta}
\affiliation{Australia Telescope National Facility, CSIRO, Space and Astronomy, PO Box 76, Epping, NSW 1710, Australia}

\author[0000-0002-8987-1544]{Akhil Jaini}
\affiliation{Centre for Astrophysics and Supercomputing, Swinburne University of Technology, Hawthorn, VIC 3122, Australia}

\author[0000-0002-6437-6176]{Clancy W. James}
\affiliation{International Centre for Radio Astronomy Research, Curtin University, Bentley, WA 6102, Australia}

\author[0000-0002-7551-2073]{John Tuthill}
\affiliation{Australia Telescope National Facility, CSIRO, Space and Astronomy, PO Box 76, Epping, NSW 1710, Australia}



\begin{abstract}

The emerging population of long-period radio transients (LPTs) show both similarities and differences with normal pulsars. 
A key difference is that their radio emission is too bright to be powered solely by rotational energy. 
Various models have been proposed (including both white-dwarf or neutron star origins), and their nature remains uncertain. 
Known LPTs have minutes to hours long spin periods, while normal pulsars have periods ranging from milliseconds to seconds. 
Here, we report the discovery of \source{}, an object with an intermediate spin period of 41 seconds, bridging the gap between LPTs and normal pulsars. 
\source{} exhibits low linear ($\sim25\%$) and circular polarisation ($\sim5\%$) and a relatively steep spectral index ($\sim-2.3$), features similar to normal pulsars. 
However, its observed spin-down properties place it below the pulsar death line, where pair production and thus radio emission are expected to cease. 
The discovery of \source{} suggests the existence of a previously undetected population within this intermediate period range, presumably missed due to selection biases in traditional pulsar search methods. 
Finding more such objects is important to fill the current gap in neutron star spin periods, improving our understanding of the relationships among rotation-powered pulsars and LPTs.

\end{abstract}

\keywords{pulsars: general --- radio continuum: general --- stars: neutron}


\section{Introduction} 
\label{sec:intro}
Radio pulsars are rapidly rotating neutron stars with periods ranging from milliseconds to seconds \citep{Manchester2005AJ....129.1993M}. 
The radio emission observed originates from a relativistic beam powered by their rotation. 
Pulsars typically exhibit moderately polarised emission, often dominated by linear polarisation (e.g., $\gtrsim40\%$ for young and energetic pulsars; \citealt{Serylak2021MNRAS.505.4483S}) and around 10--15\% circular polarisation \citep[e.g.,][]{Dai2015MNRAS.449.3223D} across the population. 
Pulsars can show a range of time-domain behaviours, including single pulse variations in shape and intensity, nulling (where pulsar emission ceases for a certain period; \citealt{Backer1970Natur.228...42B}), and scintillation (a propagation effect caused by scattering in the ionised interstellar medium for compact sources; \citealt{Johnston1998MNRAS.297..108J}). 
Pulsars are useful tools to study neutron star physics and understand their evolution path (see \citealt{Lorimer2012hpa..book.....L}). 

Recently several sources of radio pulses with periods of minutes to hours have been discovered \citep[e.g.,][]{Hurley-Walker2022Natur.601..526H,Hurley-Walker2023Natur.619..487H,Caleb2022NatAs...6..828C,Caleb2024NatAs...8.1159C}, referred to as ``long-period transients'' (LPTs). 
\add{LPTs share some properties with traditional pulsars, but their periods are so slow that the emission they produce cannot be powered by their rotation. 
In particular, their observed radio luminosity are about three orders of magnitude higher than the inferred spin-down luminosity under the assumption of magnetic dipole radiation \citep{Hurley-Walker2022Natur.601..526H,Caleb2024NatAs...8.1159C}.}
The origin of LPTs is unclear. 
White dwarfs and neutron stars have both been proposed as progenitors, but these sources do not fit canonical models of either \citep{Rea2024ApJ...961..214R}. 
Two are confirmed to originate from binary systems \citep{Hurley-Walker2024ApJ...976L..21H,deRuiter2024arXiv240811536D}, while a candidate LPT has been shown to exhibit polarisation properties consistent with a compact rotating object \citep{Dobie2024MNRAS.535..909D}. 
Revealing their origin will improve our understanding of neutron star formation processes and stellar evolution. 

One key step towards doing this is comprehensive searches for objects on intermediate timescales between LPTs (periods of minutes to hours) and pulsars (periods of milliseconds to seconds). 
Traditionally, pulsars have been discovered using time-domain data from millisecond-resolution single-dish observations, whereas almost all LPTs have been identified through image-domain searches with integration times of seconds or longer.
\add{In total power time series searches, slow variations in the total power ``baseline" can reduce sensitivity to broad pulses or long-period objects. 
In practice, these searches usually do not search for pulsars with periods longer than 10 seconds.}
Conversely, image domain searches sacrifice sensitivity to pulses shorter than the image integration time, which is generally no shorter than $\sim$10\,s. 
This leaves a sensitivity gap centered on pulses of width $\sim$1\,s, which would generally correspond to pulse periods of 10--100s.

The Australian Square Kilometre Array Pathfinder (ASKAP; \citealt{Hotan2021PASA...38....9H}) has been highly successful in finding millisecond timescale transients such as Fast Radio Bursts using the Commensal Realtime ASKAP Fast Transient (CRAFT) backend \citep{Shannon2024arXiv240802083S}. 
Recently an upgrade has been carried out to enable searches for millisecond timescale transients in visibility data --- the CRAFT Coherent upgrade \citep[CRACO;][]{Wang2025PASA...42....5W}. 
During commissioning, CRACO has so far operated with a time resolution of $\sim$110\,ms and $\sim$13.8\,ms. 
The ultimate design specifications are for $\sim$1.7\,ms resolution, with a parallel search expected to operate at 110\,ms resolution in the final design configuration.
CRACO searches for single pulses with optimally matched widths of 1 to 8 times the integration time, and hence retains excellent sensitivity to single pulses of width 1\,ms to 1\,s. 
This capability bridges the timescale gap between traditional time domain total power searches and traditional coarse time resolution imaging approaches. 

\add{In this work, we present the discovery of long period \source, which exhibits $\sim$0.5\,s duration pulses at a period of 41\,s.}
We present our observations and results in Section~\ref{sec:observations}. 
In Section~\ref{sec:discussion}, we discuss the nature of \source{} and its implication to long period objects population. 
We summarise our conclusions in Section~\ref{sec:conclusion}.





\section{Observations and Results}
\label{sec:observations}


\subsection{ASKAP}

\begin{table}
\centering
\caption{A list of observations for \source, where $F_\mathrm{cent}$ represents the observing central frequency and $T_\mathrm{obs}$ represents the total observing time. }
\begin{tabular}{ccccc}
\hline \hline
Obs. Date & Telescope & $F_\mathrm{cent}$ & $T_\mathrm{obs}$ & Obs. ID \\
(UTC) &  & (MHz) & (min) &  \\
\hline
2024~Jan~14 & ASKAP & 887.5 & 2.7 & 57299 \\
2024~Jan~18 & Parkes & 2368 & 85 & PX125 \\
2024~Jan~24 & Parkes & 2368 & 80 & PX114 \\
2024~Jan~28 & Parkes & 2368 & 80 & PX114 \\
2024~Jan~30 & ASKAP & 943.5 & 59 & 58275 \\
2024~Jan~31 & Parkes & 2368 & 105 & PX114 \\
2024~Feb~10 & Parkes & 2368 & 124 & PX114 \\
2024~Mar~08 & MeerKAT & 1284 & 39 & DDT \\
2024~Mar~08 & MeerKAT & 816 & 39 & DDT \\
2024~Apr~01 & MeerKAT & 1284 & 39 & DDT \\
2024~Apr~01 & MeerKAT & 816 & 39 & DDT \\
2024~Apr~02 & GBT & 820 & 100 & 24A-437 \\
2024~Apr~03 & GBT & 820 & 93 & 24A-437 \\
2024~Jul~03 & MeerKAT & 816 & 39 & DDT \\
2024~Jul~05 & MeerKAT & 816 & 39 & DDT \\
\hline
\end{tabular}
\label{tab:observations}
\end{table}

\begin{figure*}
    \centering
    \includegraphics[width=0.88\textwidth]{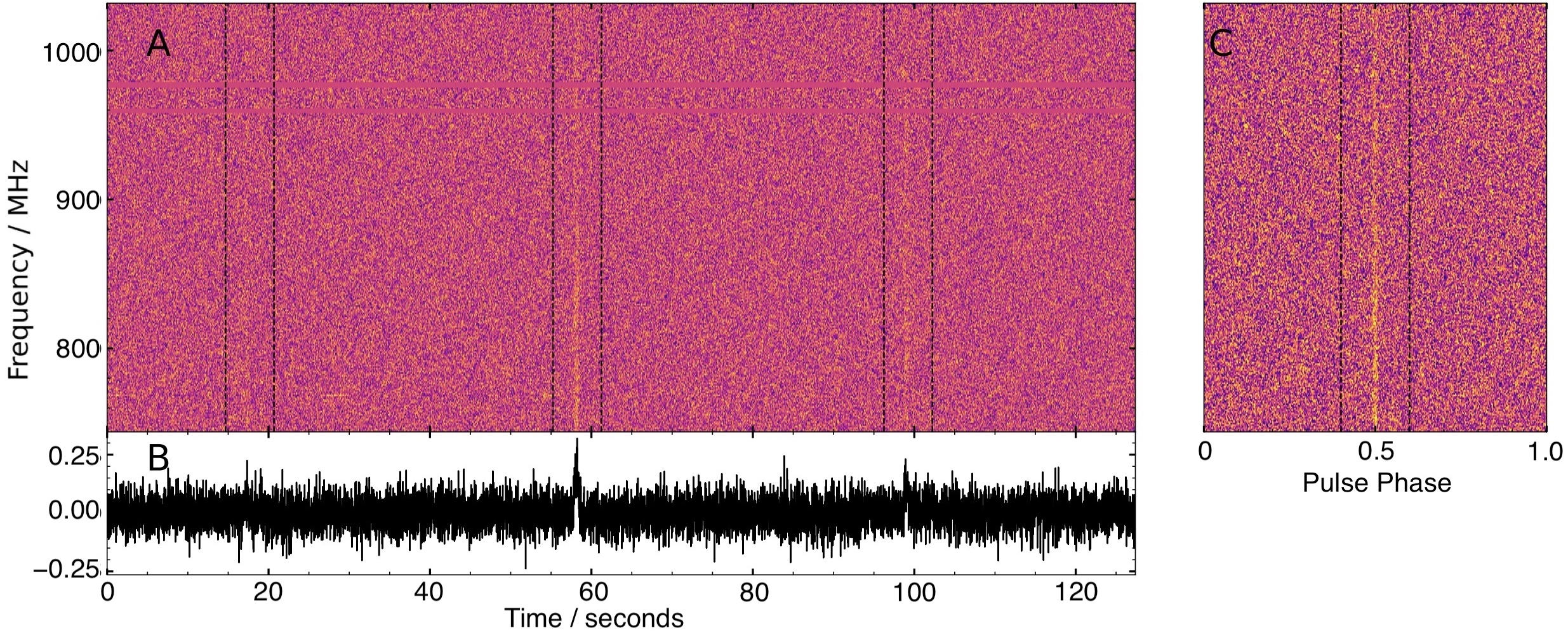}
    \caption{
    Filterbank plots for \source\ with the CRACO data. 
    \textbf{Left}: The de-dispersed tied-array beam filterbank with a time resolution of 110\,ms in the whole 2-minute discovery observation. 
    \textbf{Right}: The de-dispersed data with a time resolution of 27.6\,ms for the brightest pulse. 
}
    \label{fig:filterbank_CRACO}
\end{figure*}

We detected \source\ in an ASKAP observation on 2024~January~14, originally intended for testing the detection of interplanetary scintillation with CRACO. 
During the 2\,min test observation, the CRACO system, operating with a time resolution of 13.8\,ms, detected three consecutive pulses with pulse widths of $\sim$0.5\,second (Figure~\ref{fig:filterbank_CRACO}). 
At the time, a real-time voltage trigger for the CRACO system was not in place, and the pulses were too broad to be triggered by the incoherent-sum system \citep{Shannon2024arXiv240802083S}. 
As a result, the available data consist of 13.8-ms visibilities with one polarisation. 
The brightest pulse we detected has a signal-to-noise ratio (SNR) of $\sim$10. 
The best fit for the dispersion measure (DM) of the burst was $21\pm3$\,pc\,cm$^{-3}$. 
This is consistent with an origin in the Milky Way, with estimated distances of $\sim$0.8\,kpc using the NE2001 model \citep{Cordes2002astro.ph..7156C} and $\sim$1.1\,kpc using the YMW16 model \citep{Yao2017ApJ...835...29Y}. 
We measured a period of 40.9\,s from the three detected pulses.

The CRACO fast imaging candidate production system makes numerous approximations that affect astrometric accuracy \citep{Wang2025PASA...42....5W}. 
We therefore re-imaged the 1\,s of CRACO visibilities for the brightest pulse and obtained the pulsar position at RA $03^{\mathrm{h}}11^{\mathrm{m}}18.91^{\mathrm{s}}$, Dec $+14^{\circ}02'22.41''$ (J2000; $1\sigma$ uncertainty is approximately $2''$ in both direction). 
We measured a flux density of $30.9\pm2.3$\,mJy for the brightest pulse in the 1\,s CRACO image.

We performed a 1-hour follow-up observation using ASKAP on 2024~January~30. 
The real-time single pulse search carried out automatically with the CRACO system did not result in any detections, but offline analysis of the data revealed faint pulsed emission (SNR $\approx5$) after folding the data to the 40.9\,s period. 
We then carried out further follow-up observations with Parkes, MeerKAT and the Green Bank Telescope. 
These observations are listed in Table~\ref{tab:observations}, and we describe the specifications and results below.

\subsection{Parkes/Murriyang}

We performed five follow-up observations using the Parkes ultrawide band low (UWL) receiver \cite[][]{Hobbs2020PASA...37...12H}. 
We used a time resolution of 64\,$\mu s$ and a frequency resolution of 0.5\,MHz for the first epoch, and a coarser temporal and spectral resolution of 1024\,$\mu s$ and 0.5\,MHz for the subsequent four epochs. 
We searched for single pulses using a HEIMDALL-FETCH based multi-tiered sub-band search pipeline \citep{Barsdell2012MNRAS.422..379B}, which did not result in any detection. 
We detected pulsed emission from \source\ in the second epoch (the observation on 2024~January~24) after folding the data, confirming the 40.9\,s period measured by ASKAP. 
The other epochs did not result in any detection after folding. 
The non-detection is possibly caused by a combined effect of Parkes RFI environment, baseline variations, and scintillation due to the low DM and high galactic latitude of \source.  

\begin{figure*}
    \centering
    \includegraphics[width=\textwidth]{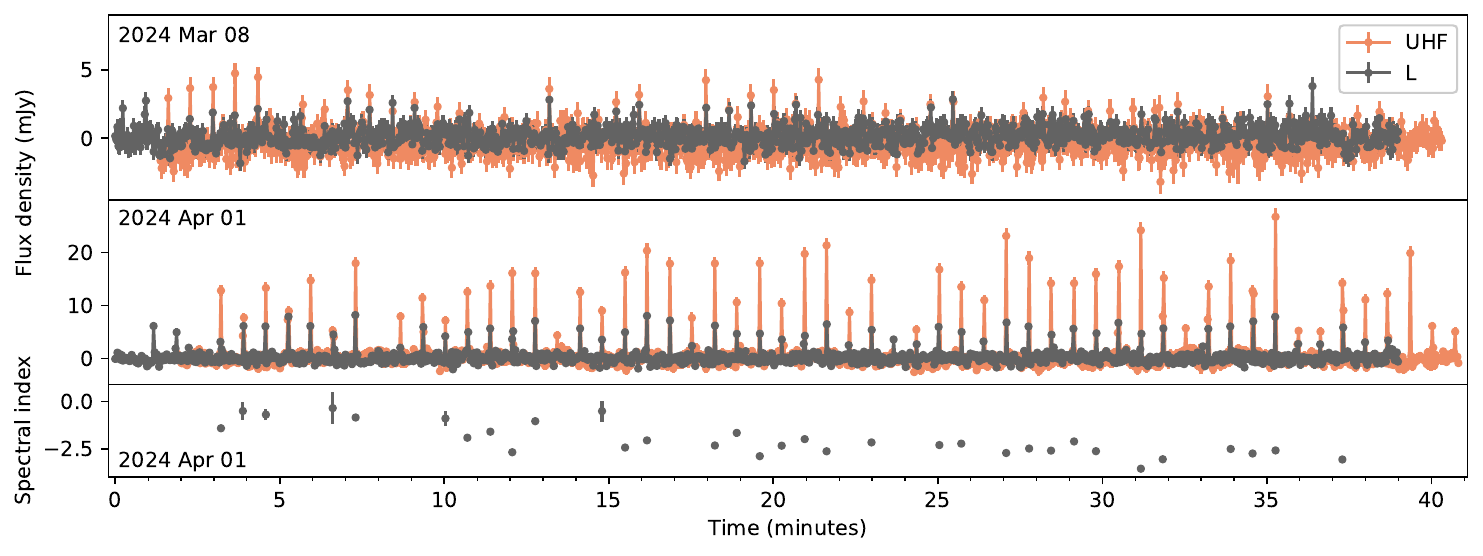}
    \caption{
    Radio lightcurve of \source{} extracted from 2-second images from the first two MeerKAT observations. The orange points represent the flux density measured at the UHF band, while the grey points represent the flux density measured at the L band. The bottom panel displays the spectral indices (with errors) calculated from the UHF and L band data during the MeerKAT epoch 2 observation.
}
    \label{fig:meerkat_lightcurve}
\end{figure*}

\subsection{MeerKAT}

We observed \source\ in four epochs with MeerKAT using both pulsar mode (Pulsar Timing User Supplied Equipment; PTUSE) and imaging mode simultaneously. 
The first two epochs were observed at UHF (816\,MHz) and L (1284\,MHz) band simultaneously via sub-arraying, with a 2\,s integration time for imaging mode and 60\,$\mu s$ sampling time for pulsar mode. The remaining two epochs were observed solely at the UHF band with an 8\,s integration time for imaging mode and 60\,$\mu s$ sampling time for pulsar mode. 

We used PKS~B1934$-$638 as the primary bandpass and flux calibrator, and J0318$+$1628 as the secondary phase calibrator. 
We flagged, calibrated, and imaged visibilities using \code{oxkat} \citep{Heywood2020ascl.soft09003H}. 
For the first two epochs, we created 2\,s snapshot images using \code{wsclean \citep{Offringa2014MNRAS.444..606O}}. 
\source\ has been detected every $\sim$20 2-s snapshot images, and the lightcurve is shown in Figure~\ref{fig:meerkat_lightcurve}. 
\add{We note that \source{} exhibits clear variations in single-pulse amplitudes. }
We are able to obtain an improved position at RA $03^{\mathrm{h}}11^{\mathrm{m}}18.59^{\mathrm{s}}\pm0.05''$, DEC $+14^{\circ}02'20.49''\pm0.05''$ (J2000; errors are statistical) using the 2\,s snapshot image with the brightest detection. 
We measured a median spectral index of $-2.3\pm0.2$  based on the UHF and L band images in the second MeerKAT observing epoch. 
The spectral index varies from $-0.5$ to $-3.5$, presumably due to scintillation. 
We combined all 2-s snapshot images in off-pulse phase, and measured an off-pulse emission $3\sigma$ limit of $<40\,\mu$Jy. 

The PTUSE data were processed using \code{psrchive}. 
We are able to measure a more precise DM of $19.9\pm1.4$\,pc\,cm$^{-3}$. 
Figure~\ref{fig:pulse_profile} shows the folded pulse profiles at different epochs. 
\add{\source{} displays a clear double-peak structure, with a potential triple-peak feature appearing in the epoch 4 observation.}
\add{We observed a bright narrow-band scintillation pattern in the folded pulses from epochs 3 (around 550\,MHz to 650\,MHz) and epoch 4 (around 550\,MHz to 600\,MHz). 
We measured the scintillation bandwidth to be $26\pm5$\,MHz using \code{scintools} \citep{Reardon2020ascl.soft11019R} from the epoch 4 observation.}
\add{We also predicted the scintillation timescale at 800\,MHz to be approximately 10\,minutes for diffractive scintillation and 4\,days for refractive scintillation, based on the NE2001 and YMW16 model. 
The predicted refractive modulation index is about 0.4, consistent with what we observed from epoch-to-epoch variations. 
The decorrelation bandwidth predicted by the NE2001 and YMW16 model is about 2\,MHz, smaller than the measured 26\,MHz scintillation bandwidth, which may be due to variations in model predictions at this latitude. }
\add{We measured a low rotation measure (RM) consistent with 0\,rad\,m$^{-2}$. }
\add{The polarisation position angle (PA) undergoes a $90^\circ$ jump at MeerKAT epoch 3 and 4, and similar jumps have been observed in some pulsar pulses \citep[e.g.,][]{Gould1998MNRAS.301..235G}.}
We obtained a pulse width $W_{50}$ (full-width at half-maximum) of $432\pm6$\,ms from the folded pulse profile at UHF band. 
The PTUSE data have significant contamination from time-varying RFI, making it hard to identify any single pulse sub-structure.

\subsection{Green Bank Telescope}

\begin{figure*}
    \centering
    \includegraphics[width=0.88\textwidth]{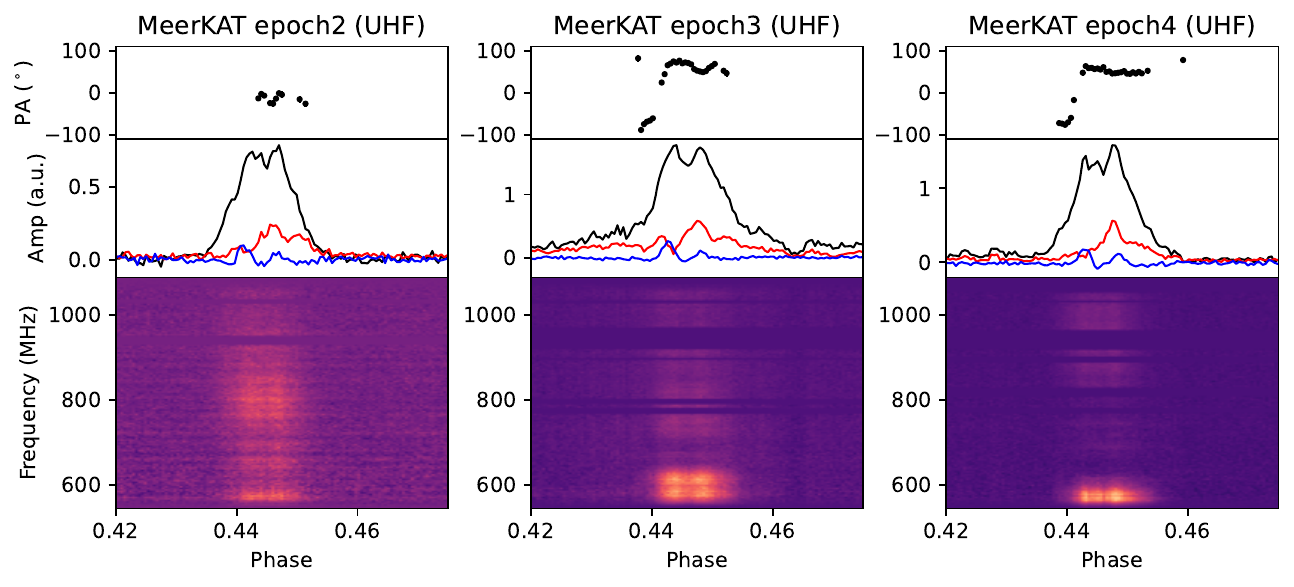}
    \includegraphics[width=0.62\textwidth]{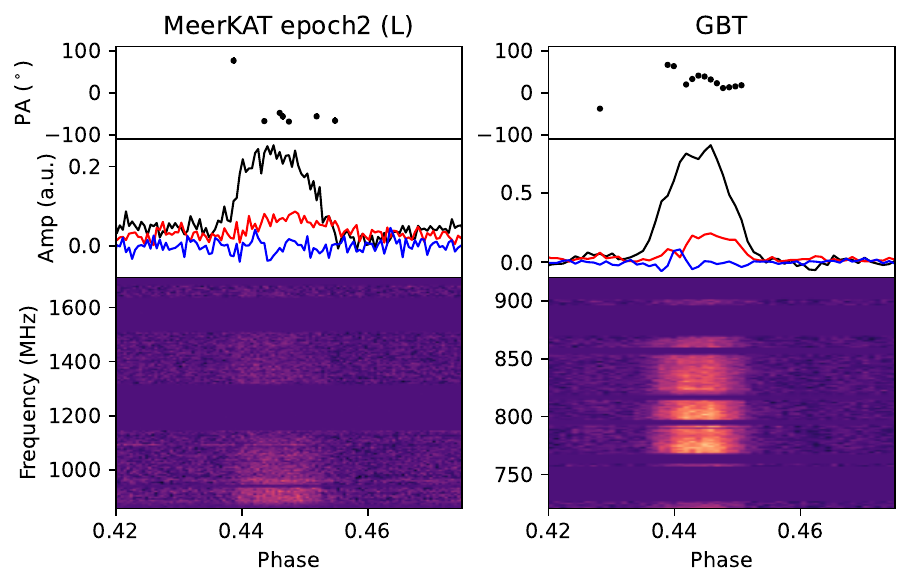}
    \caption{
    A sample of folded radio pulses from \source{}. For each observation, the top panel shows the linear polarisation position angle (PA) for detections of linear polarisation greater than three times the off-pulse noise. 
    The middle panel presents the average intensity versus phase: the total intensity is shown as a black solid line, linear polarisation is shown as the red solid line, and circular polarisation as the blue solid line. 
    The bottom panel shows the pulse dynamic spectrum. 
    `a.u' stands for arbitrary units.
    \add{The dark horizontal regions represent flagged data due to RFI.}
    \add{The number of averaged pulses are 58 for the MeerKAT observations and 136 for the GBT observation.}
    \add{The PAs in MeerKAT epoch 2 are different from those in other epochs, possibly due to low signal-to-noise ratio and inadequate number of averaged pulses. }
}
    \label{fig:pulse_profile}
\end{figure*}

We observed \source\ with the Green Bank radio telescope (GBT; project ID GBT24A$-$437) in pulsar search mode from the prime focus 800\,MHz receiver with a bandwidth of 200\,MHz, at a central frequency of 820.4\,MHz. 
The source was observed for a total of 4 hours, with each observing session spanning approximately 2 hours. 
The lower levels of RFI at the Green Bank site, compared to Parkes, reduce the level of baseline variations, which is important for long-period objects detection. 
We were able to detect \source\ in both observing sessions (Figure~\ref{fig:pulse_profile}).

\subsection{Pulsar timing}

We conducted timing analysis using standard techniques, using observations obtained with the GBT, MeerKAT, and Parkes.
We formed arrival times using the \textsc{pat} tool in  \textsc{psrchive} \cite[][]{Hotan2004PASA...21..302H}.
We developed an initial pulsar ephemeris using \textsc{tempo2} \cite[][]{2006MNRAS.369..655H}.
In the timing model, the pulsar position was fixed at the value derived from MeerKAT imaging, so we fit only for the pulsar spin frequency and frequency derivative.
To place a robust constraint on the frequency derivative we undertook Bayesian analysis using the \textsc{temponest} package \cite[][]{2014MNRAS.437.3004L}.
We sample both the spin frequency and frequency derivative. We searched for two modifications to the formal arrival time uncertainties. 
The modelled arrival times are $\sigma = \sqrt{F^2 \sigma_i^2 + \sigma_Q^2}$, where $\sigma_i$ is the original uncertainties, $F$ is a scaling factor that modifies the orignal uncertainties and $\sigma_Q$ is an additional quadarature uncertainty.  
These uncertainties account for pulse jitter \cite[][]{2010arXiv1010.3785C} and distortions to the pulse profile from radio frequency interference. 
We constrain the spin frequency derivative to be to be $|\dot{\nu}|< 7.47\times10^{-18}$\,s$^{-2}$. This correponds to $|\dot{P}|< 1.25\times10^{-14}$\, s\,s$^{-1}$.
The properties of \source{}, including measured parameters and derived parameters from the timing analysis, are listed in Table~\ref{tab:pulsar}. 

\begin{table}
\centering
\caption{Timing and model parameters for \source, including measured parameters and derived parameters from the timing analysis. }
\begin{tabular}{ll}
\hline \hline
Measured Parameters & Values \\
\hline
Right ascension (J2000) & $03^{\mathrm{h}}11^{\mathrm{m}}18.59^{\mathrm{s}}\pm0.05''$  \\
Declination (J2000) & $+14^{\circ}02'20.49''\pm0.05''$ \\
Spin period & $40.9106968\pm(2\times10^{-7})$\,s\\
Spin period derivative & $<1.25\times10^{-14}$\, s\,s$^{-1}$ \\
Timing span (MJD) & 60323.6 to 60496.2 \\
Number of TOAs & 20 \\
Dispersion measure, DM & $19.9\pm1.4$\,pc\,cm$^{-3}$ \\
$W_{50}$ (UHF band) & $432\pm6$\,ms \\
$W_{50}$ (L band) & $456\pm14$\,ms \\
Spectral index (median) & $-2.3\pm0.2$ \\
Linear polarisation & 22\%--26\%\\
Circular polarisation & 4\%--6\% \\
\hline
Derived Parameters & \\
\hline
Galactic longtitude & $+166^{\circ}48'02.38''$ \\
Galactic latitude & $-36^{\circ}38'59.33''$\\
Distance (YMW16) & \add{971\,pc} \\
Distance (NE2001) & \add{741\,pc} \\
Chracteristic age & $> 51$ Myr \\
Surface dipole magnetic field & $< 2 \times 10^{13}\,$G\\
strength &  \\
Spin-down luminosity & $< 1 \times 10^{28}\,$ erg s$^{-1}$\\
\hline
\end{tabular}
\label{tab:pulsar}
\end{table}

\subsection{Archival data}

We investigated archival radio and multi-wavelength data that covered the location of \source. 
The only ASKAP survey covering the source field is the Rapid ASKAP Continuum Survey (RACS; \citealt{McConnell2020PASA...37...48M}).  
RACS has multiple epochs at different frequencies, with this position being covered five times between 2019 April and 2023 December \citep{Hale2021PASA...38...58H,Duchesne2023PASA...40...34D}. 
No detections were made at the source location, with 5$\sigma$ upper limits of 1.7\,mJy\,beam$^{-1}$ at 887.5\,MHz, 1.1\,mJy\,beam$^{-1}$ at 943.5\,MHz and 1655.5\,MHz, and 0.8\,mJy\,beam$^{-1}$ at 1367.5\,MHz. 
These non-detections can be explained by its low duty cycle. 
There are no detections in other archival radio continuum surveys, including the GaLactic and Extragalactic All-sky Murchison Widefield Array (GLEAM) survey at 200\,MHz \citep{Hurley-Walker2017MNRAS.464.1146H} and the Karl G. Jansky Very Large Array Sky Survey (VLASS) at 3\,GHz \citep{Lacy2020PASP..132c5001L}.

There is no Gaia \citep{GaiaCollaboration2023A&A...674A...1G}, WISE \citep{Wright2010AJ....140.1868W}, 2MASS \citep{Skrutskie2006AJ....131.1163S}, and Fermi \citep{Acero2015ApJS..218...23A} source within the localisation error box. 
From the 2MASS survey, we estimated the deepest $3\sigma$ limits on the $J$- and $K$-band magnitude of 17.5\,mag (Vega) and 16.0\,mag (Vega), respectively. 
We retrieved the typical color and absolute magnitude for main-sequence and ultra-cool stars from \citep{Pecaut2013ApJS..208....9P}, and used \textsc{dustmaps} \citep{Green2018JOSS....3..695G} to estimate the extinction. We can rule out any stellar object with stellar type earlier than M4 at a distance of 1.1\,kpc.

\section{Discussion}
\label{sec:discussion}

\begin{figure*}
    \centering
    \includegraphics[width=0.75\textwidth]{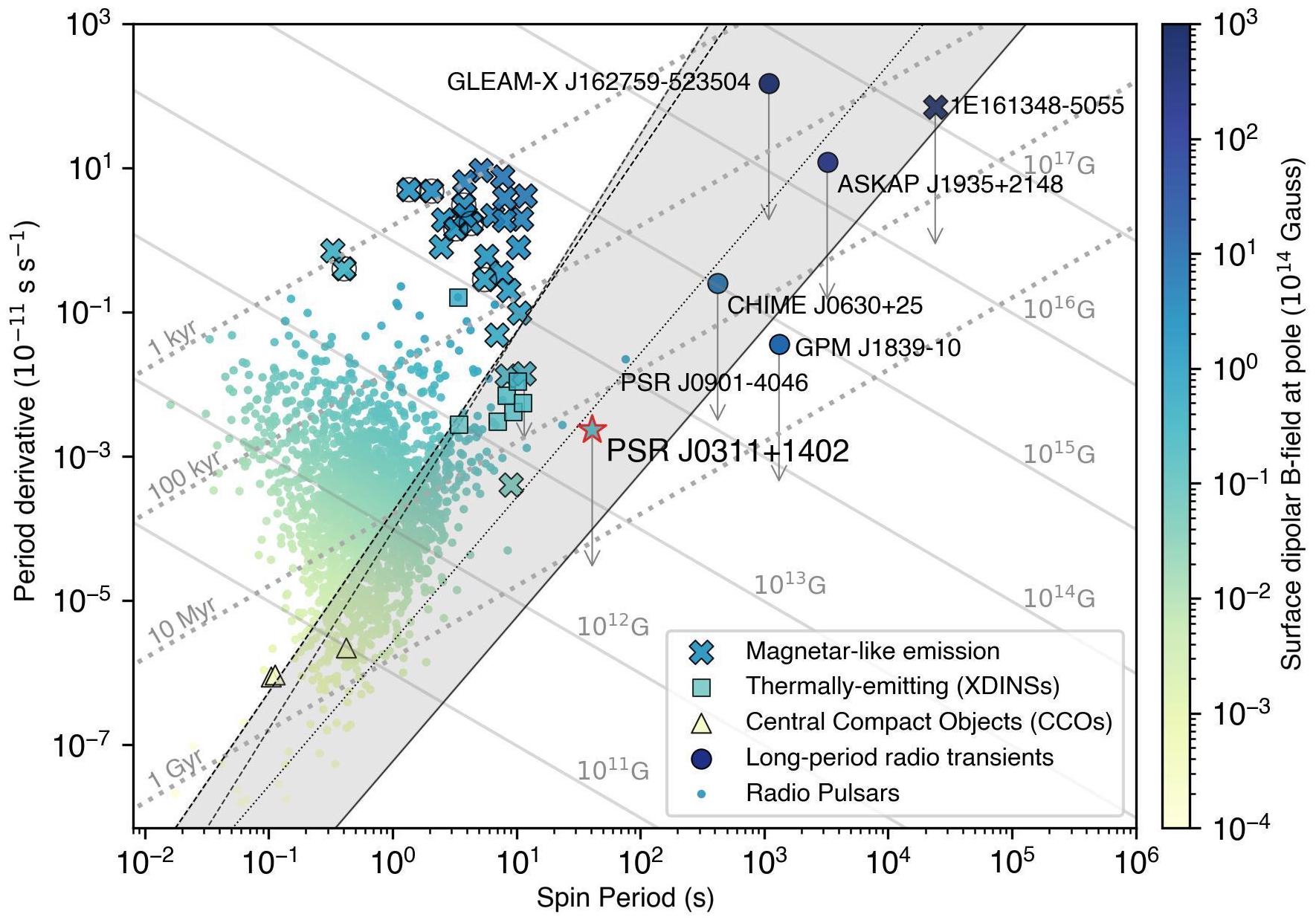}
    \caption{Diagram of spin period versus period derivatives for pulsars and LPTs. 
    The red star marks \source{}. Dots represent radio pulsars from the ATNF pulsar catalogue \citep{Manchester2005AJ....129.1993M}, and circles represent recently discovered LPTs \citep{Hurley-Walker2022Natur.601..526H,Hurley-Walker2023Natur.619..487H,Caleb2022NatAs...6..828C,Caleb2024NatAs...8.1159C}.
    Markers are colored by surface dipolar magnetic field at the pole, assuming spin-down due to dipolar losses. 
    Dashed lines correspond to theoretical death lines for a pure dipole \citep{Chen1993ApJ...402..264C,Zhang2000ApJ...531L.135Z}, dotted lines for a twisted dipole \citep{Chen1993ApJ...402..264C,Zhang2000ApJ...531L.135Z}, and solid lines for a twisted multipole configuration \citep{Chen1993ApJ...402..264C}. 
    Figure adapted from \citet{Hurley-Walker2023Natur.619..487H}. }
    \label{fig:ppdot}
\end{figure*}

We note that \source{} shares many characteristics with normal pulsars, such as a steep spectral index and similar pulse morphology. 
However, its period is significantly longer than that of the typical pulsar population. 
\add{To determine whether \source{} is part of the normal pulsar population, one approach is to check whether its radio luminosity can be powered solely by its rotational energy. }
This is particularly relevant given that the currently known LPTs exhibit much higher radio luminosities, suggesting a different emission mechanism.

\subsection{Radio luminosity}


The radio luminosity $L$ of \source{} can be calculated by 
\begin{equation}
    L = 4\pi \sin^2\left(\frac{\rho}{2}\right)d^2 \int S(f)_{\rm peak}\, \mathrm{d}f
\end{equation}
where $\rho$ is the beam opening angle, $d$ is the distance, $S(f)_{\rm peak}$ is the maximum intensity of the pulse profile, and $f$ describe the frequency of the measurement \citep{Lorimer2012hpa..book.....L}. 
Based on the MeerKAT UHF-band observation we estimated $S_{\rm peak} \approx 30$\,mJy at $f=816$\,MHz with a median spectral index of $-2.3$.

For radio emission to occur in a rotationally powered pulsar, the polar cap radius, $R_p$, must exceed the curvature photon gap height, $\ell_{\rm gap}$, near the surface \citep{1979ApJ...231..854A,1981ApJ...248.1099A,2023MNRAS.520.1872B}. The polar cap radius of \source{} is given by 
\begin{equation}
    R_p \simeq \sqrt{\frac{2\pi R^3}{cP}} \sim 2\times10^3 \, \rm cm,
    \label{eq:PC}
\end{equation}
where $R\simeq10^6 \,$cm is the assumed radius of the neutron star. We estimated $\ell_{\rm gap}$ using the equation from \citet{2015ApJ...810..144T} and neglected order unity factors,
\begin{equation}
\begin{split}
    \ell_{\rm gap} &\simeq 2 \times 10^4 \left(\frac{\rho_c}{10^7\,{\rm cm}} \right)^{2/7} \left(\frac{P}{1\,{\rm s}}\right)^{3/7} \left(\frac{B}{10^{12}\,{\rm G}}\right)^{-4/7} \\
    &\gtrsim 10^4\, {\rm cm},
\end{split}
\end{equation}
where $\rho_c= 10^7\,$cm is the typical radius of curvature and $B=2\times 10^{13}$\,G is the estimated surface magnetic field strength. 
\add{Given the possible variation in radius of curvature for a 41-s pulsar, we also estimated $\ell_{\rm gap}$ using $\rho_c$ values from $10^6\,$cm to $10^8\,$cm. The estimated lower limits of $\ell_{\rm gap}$ remain consistent at the order of $10^4$\,cm. }
The discrepancy between $\ell_{\rm gap}$ and $R_p$ may arise because the expression for the polar cap radius might not be applicable to pulsars with very long periods. 
We assume the polar cap radius of our source $R_p$ to be at least as large as the gap height $\ell_{\rm gap}$, i.e., $R_p = 10^4\, {\rm cm}$.

The polar cap radius is related to the emission height through
\begin{equation}
    \frac{\sin^2\theta_{\rm em}}{r_{\rm em}} = \frac{\sin^2\theta_{p}}{R},
\end{equation}
where ($r_{\rm em}, \theta_{\rm em}$) are the coordinates of the emission point and $\sin \theta_{p} = R_p / R$. The emission height can be expressed as 
\begin{equation}
\begin{split}
     r_{\rm em} &\simeq 4 \times 10^7 \left ( \frac{f_0}{1 \,\rm GHz} \right)^{\beta} \left( \frac{\dot{P}}{10^{-15} \,\rm s\, s^{-1}} \right)^{0.07} \left( \frac{P}{\rm s} \right)^{0.3} \\
     &\sim 10^8 \, {\rm cm},
\end{split}
\end{equation}
where $\beta = 0.26$ is a typical value for radio pulsars \citep{Kijak2003A&A...397..969K}. 
\add{The estimated $r_{\rm em}$ is much smaller than the light cylinder radius $r_{\rm LC}\sim10^{11}$\,cm.  
If we instead assume a larger $r_{\rm em}\sim 0.01r_{\rm LC}$, the estimated beam opening angle $\rho$ would be about $35^\circ$, which is significantly larger than the observed pulse width $W\sim4^\circ$. 
Therefore, we consider $r_{\rm em}\sim10^8$\,cm to be a reasonable estimate.}
We then have $\theta_{em} = 0.13$, and the beam opening angle can be approximated as $\rho = 3\theta_{em}/2 \approx 0.20$. 
Alternatively, using an empirical relationship where $\rho \propto P^{-1/2}$ and assuming a typical $\rho \approx 6^\circ$ for a $P=1$\,s pulsar \citep{Gould1994PhDT.......164G}, the estimated $\rho$ for a 40-s pulsar would be approximately 0.02. 

The radio spectrum of pulsars often peak and turn over at relatively low radio frequencies ($\sim$100\,MHz; \citealt{Lorimer2012hpa..book.....L}). 
If we assume the majority of the flux density is contributed from the frequency range between $f_1 = 100$\,MHz and $f_2 = 10$\,GHz, the luminosity of \source{} is estimated to be $\sim10^{28}$\,erg\,s$^{-1}$ (for $\rho\approx0.2$) and $\sim10^{26}$\,erg\,s$^{-1}$ (for $\rho\approx0.02$). 
This is comparable to or less than the estimated upper limit for the spin-down luminosity, which is $\sim10^{28}$\,erg\,s$^{-1}$. 
Given that \source\ is a variable source and lacks measurements at low ($<500$\,MHz) and high ($>2$\,GHz) frequency, the uncertainty of this estimation can be large.   
However, based on current measurements, we cannot rule out the possibility that \source{} is a rotation-powered pulsar. 

\subsection{Pulsar death line}

Although the pulsar emission mechanism is not yet fully understood, it is commonly thought to be connected with ubiquitous electron-positron pair production in the magnetosphere \citep[e.g.,][]{Philippov2022ARA&A..60..495P}.
As a pulsar ages and spins more slowly, the particle acceleration potential is expected to be lower and the coherent radio emission will eventually cease. 
Various theoretical models predict the conditions under which pair production can no longer be produced, defining what is known as the pulsar death lines \citep{Chen1993ApJ...402..264C, Zhang2000ApJ...531L.135Z}. 
Observationally, the spin-down properties of known pulsars are generally consistent with the boundaries defined by these death lines.

Figure~\ref{fig:ppdot} shows the radio transient population in the period-period derivative plot with several proposed death lines. 
We note that \source{} lies apart from the general pulsar population, located below the death lines for neutron stars with a pure or twisted dipole magnetic field (case 1 and 2 in \citealt{Chen1993ApJ...402..264C}). 
\source{} is not yet confirmed to lie below the death line for the extreme case where the polar cap is so twisted that curvature radiation photons are emitted nearly parallel to the magnetic field (case 4 in \citealt{Chen1993ApJ...402..264C}). 
Ongoing timing observations would provide a more precise measurement in the period-period derivative parameter space, offering deeper insights into its evolutionary state. 

\subsection{Relation to LPTs}

\begin{figure}
    \centering
    \includegraphics[width=\columnwidth]{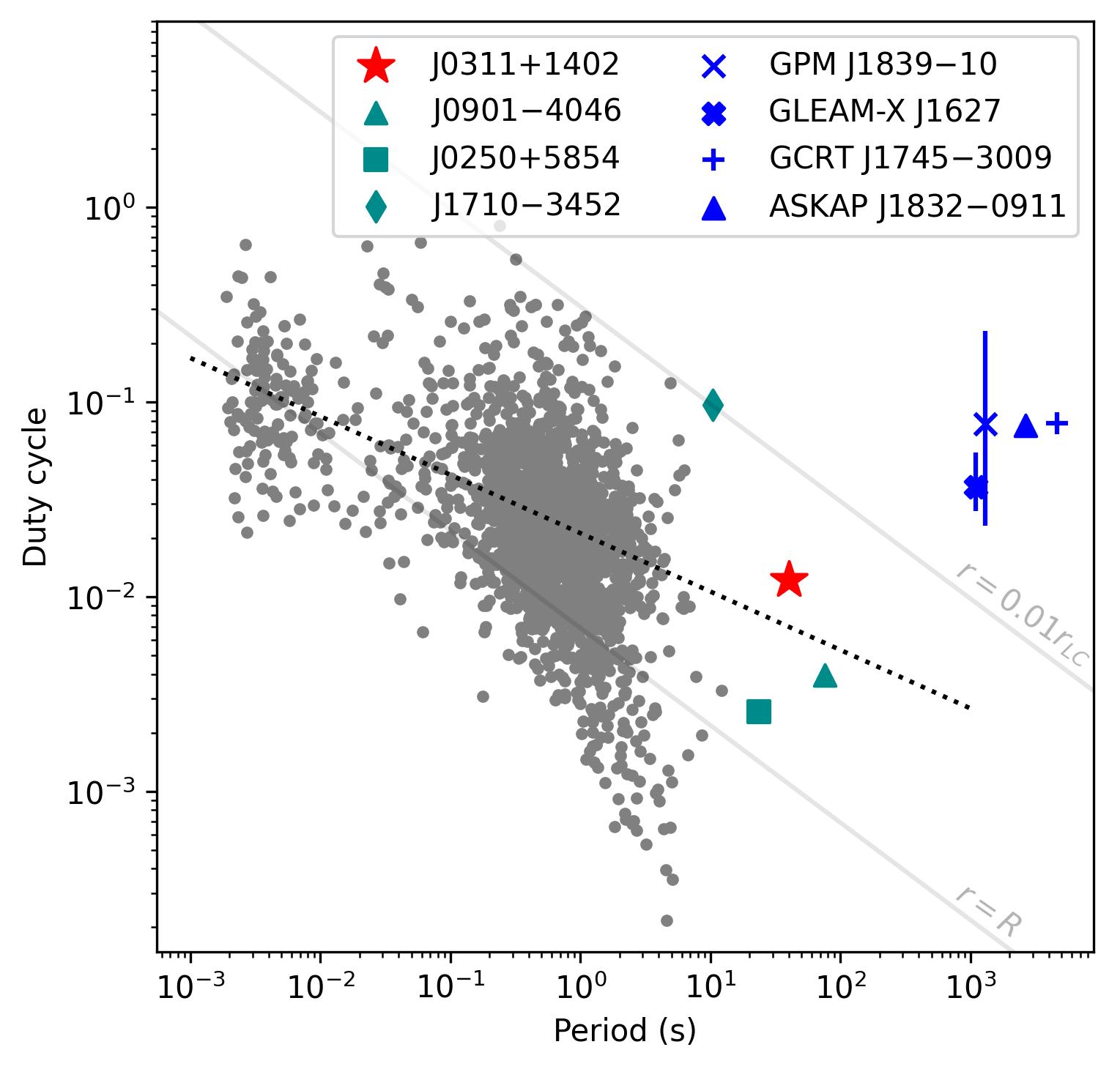}
    \caption{Pulsar period versus duty cycle. The dots represent pulsars from the ATNF pulsar catalog \citep{Manchester2005AJ....129.1993M}, while the dotted line shows the empirical relationship between pulsar spin period and pulse profile width \citep{Johnston2019MNRAS.485..640J}. 
    \add{The grey lines represent the theoretical relationship between pulsar period and pulse width, assuming an inclination angle of $45^\circ$ and an impact parameter of $0^\circ$. The two lines correspond to emission heights of $R$ (the neutron star radius) and $0.01r_{\rm LC}$, where $r_{\rm LC}$ is the light cylinder radius.}
    \source{} is marked with a red star, and known LPTs from literatures are also included \citep{Hyman2005Natur.434...50H,Tan2018ApJ...866...54T,Surnis2023MNRAS.526L.143S,Caleb2022NatAs...6..828C,Caleb2024NatAs...8.1159C,Hurley-Walker2022Natur.601..526H,Hurley-Walker2024ApJ...976L..21H, 2024arXiv241116606W}.}
    \label{fig:pulsar_duty_cycle}
\end{figure}

LPTs have periods on the order of minutes to hours, which are significantly longer than typical radio pulsars. 
They can be broadly divided into two types, isolated or in a binary system. 
There are currently four instances of LPTs that are in a binary system, with a period ranging from $\sim$2\,minutes to $\sim$2\,hours \citep{Marsh2016Natur.537..374M,Pelisoli2023NatAs...7..931P,Hurley-Walker2024ApJ...976L..21H,deRuiter2024arXiv240811536D}. 
These systems are suggested to be composed of a white dwarf and an M-dwarf star. 
In the cases of \arsco{} and \pelisoli{}, the pulsation period arises from the beat frequency between the spin period of the white dwarf and the orbital period of the companion star.
Optical and near-IR counterparts are common in these systems due to the M dwarf companion, while ultra-violet and X-ray emissions may arise from the synchrotron radiation of relativistic electrons powered by the magnetic field of the white dwarf. 
Given the absence of apparent counterparts in the electromagnetic spectrum outside of the radio frequency, we suggest that \source{} is less likely to be a radio-pulsating white dwarf binary.

The nature of isolated LPTs remains more uncertain, with possible candidates including white dwarfs, regular radio pulsars, and magnetars \citep[e.g.,][]{Rea2024ApJ...961..214R}. 
To date, only a few apparently isolated LPTs have been found \citep[e.g.,][]{Hurley-Walker2022Natur.601..526H,Hurley-Walker2023Natur.619..487H,Caleb2024NatAs...8.1159C}. 
These objects typically exhibit a high degree of linear polarization, often exceeding 80\%, and some show significant circular polarisation $\gtrsim 40\%$. 
Additionally, their radio luminosities are comparable to, or even surpass, their spin-down luminosities, suggesting that the radio emission is not solely powered by the compact object's spin-down. 
\add{These LPTs are all located in the inner Galactic plane, at Galactic latitudes of $|\ell| < 10^{\circ}$. }
These objects exhibit a wide range of spectral indices, from a positive spectral index of $+0.4$ in \askapulp{} to a steep spectral index of $-3.2$ observed in \gpm{} \citep{Hurley-Walker2023Natur.619..487H,Caleb2024NatAs...8.1159C}.
In contrast, \source{} consistently exhibits relatively low linear polarization ($\sim$25\%) and circular polarization ($\sim$5\%) across all observed epochs, and is located at a high Galactic latitude $\sim-36^\circ$. 
These properties are distinct from known LPTs. 

We note that the duty cycle for \source{} is very low, which is not unexpected as the pulsar beam radius scales as $\rho\propto P^{-1/2}$. 
The observed pulse width, $W$, depends not only on the pulsar beam radius but also on the emission height, viewing angle, and the geometry of the star. 
The pulse width is therefore one of the simplest indicators of radio emission physics of the neutron star. 
\citet{Johnston2019MNRAS.485..640J} examined a homogenous sample of $600$ pulsars observed with Parkes and found that the observed profile width, $W$, follows a scaling relationship of $W\propto P^{-0.3}$. 
As shown in Figure~\ref{fig:pulsar_duty_cycle}, \source, along with \lofarulp{} \citep{Tan2018ApJ...866...54T} and \meerkatulp{} \citep{Caleb2022NatAs...6..828C}, is consistent with this trend. 
In contrast, other LPTs, including \gleamx{} \citep{Hurley-Walker2022Natur.601..526H}, \gpm{} \citep{Hurley-Walker2023Natur.619..487H}, and \andyulp{} \citep{2024arXiv241116606W}, are located far away from the scaling relationship. 
Although the analysis is limited by small-number statistics, this distinction further suggests that \source{} is more likely to be a pulsar rather than an LPT.

\section{Conclusion}
\label{sec:conclusion}

The discovery of \source{} with ASKAP provides new insights into the relationship between neutron stars and LPTs. 
With its intermediate spin period of 41 seconds, \source{} sits squarely in the previously unpopulated parameter space between traditional pulsars and LPTs. 
Its properties, including a relatively plausible radio luminosity, steep spectral index, and low polarization fractions, strongly suggest it is a pulsar. 
However, its position below the pulsar death line in the $P-\dot{P}$ diagram raises questions about its emission mechanism. 
Further observations and timing studies of \source{} could refine its spin-down properties and improve our understanding of its evolution. 
This discovery suggests there may be a previously undetected population of neutron stars with intermediate spin periods, likely missed due to limitations in traditional search methods.
The ongoing development and operation of the ASKAP CRACO system will likely enable the discovery of more such objects, contributing to a better understanding of the neutron star population and their diverse emission processes.


\section{Acknowledgements}
We thank the MeerKAT, Parkes, and GBT directors for approving our DDT observations, and S. Buchner for assistance in scheduling and conducting MeerKAT observations.
We thank Simon Johnston and Matthew Bailes for valuable discussions. 
We thank the referee for their helpful comments and suggestions. 
YW and JJ-S acknowledge the support of the Australian Research Council grant DP220102305. 
YW and RMS acknowledge support through ARC Future Fellowship FT190100155.
MG and CWJ acknowledge support from the Australian Government through
the Australian Research Council's Discovery Projects
funding scheme (DP210102103). MG acknowledges support through UK STFC Grant ST/Y001117/1 and support from the Inter-University Institute for Data Intensive Astronomy (IDIA). IDIA is a partnership of the University of Cape Town, the University of Pretoria and the University of the Western Cape. For the purpose of open access, the author has applied a Creative Commons Attribution (CC BY) licence to any Author Accepted Manuscript version arising from this submission.
This work was performed on the OzSTAR national facility at Swinburne University of Technology. The OzSTAR program receives funding in part from the Astronomy National Collaborative Research Infrastructure Strategy (NCRIS) allocation provided by the Australian Government, and from the Victorian Higher Education State Investment Fund (VHESIF) provided by the Victorian Government. 
This scientific work uses data obtained from Inyarrimanha Ilgari Bundara / the Murchison Radio-astronomy Observatory. We acknowledge the Wajarri Yamaji People as the Traditional Owners and native title holders of the Observatory site. CSIRO's ASKAP radio telescope is part of the Australia Telescope National Facility (\url{https://ror.org/05qajvd42}). Operation of ASKAP is funded by the Australian Government with support from the National Collaborative Research Infrastructure Strategy. ASKAP uses the resources of the Pawsey Supercomputing Research Centre. Establishment of ASKAP, Inyarrimanha Ilgari Bundara, the CSIRO Murchison Radio-astronomy Observatory and the Pawsey Supercomputing Research Centre are initiatives of the Australian Government, with support from the Government of Western Australia and the Science and Industry Endowment Fund.
Murriyang, the Parkes radio telescope, is part of the Australia Telescope National Facility (\url{https://ror.org/05qajvd42}) which is funded by the Australian Government for operation as a National Facility managed by CSIRO.
We acknowledge the Wiradjuri people as the Traditional Owners of the Observatory site.
The MeerKAT telescope is operated by the South African Radio Astronomy Observatory, which is a facility of the National Research Foundation, an agency of the Department of Science and Innovation.
Observations made use of the Pulsar Timing User Supplied Equipment (PTUSE) servers at MeerKAT which were funded by the MeerTime Collaboration members ASTRON, AUT, CSIRO, ICRAR-Curtin, MPIfR, INAF, NRAO, Swinburne University of Technology, the University of Oxford, UBC and the University of Manchester.  The system design and integration was led by Swinburne University of Technology and Auckland University of Technology in collaboration with SARAO and supported by the ARC Centre of Excellence for Gravitational Wave Discovery (OzGrav) under grant CE170100004.
This research has made use of the VizieR catalogue access tool, CDS, Strasbourg, France \citep{10.26093/cds/vizier}. 
The original description of the VizieR service was published in \citet{Ochsenbein2000A&AS..143...23O}. 

%

\vspace{5mm}
\facilities{ASKAP, Parkes, GBT, MeerKAT}


\software{
aplpy~\citep{Robitaille2012ascl.soft08017R}, 
astropy~\citep{AstropyCollaboration2013A&A...558A..33A,AstropyCollaboration2018AJ....156..123A}, 
astroquery~\citep{Ginsburg2019AJ....157...98G}, 
dspsr~\citep{vanStraten2011PASA...28....1V}, 
psrchive~\citep{Hotan2004PASA...21..302H}, 
matplotlib~\citep{Hunter2007CSE.....9...90H},
numpy~\citep{Harris2020Natur.585..357H}, 
oxkat\footnote{\url{https://github.com/IanHeywood/oxkat}}~\citep{Heywood2020ascl.soft09003H}, 
scintools\footnote{\url{https://github.com/danielreardon/scintools}}~\citep{Rea2024ApJ...961..214R}, 
scipy~\citep{Virtanen2020NatMe..17..261V}, 
and wsclean~\citep{Offringa2014MNRAS.444..606O}. 
}







\bibliography{citation}{}
\bibliographystyle{aasjournal}



\end{document}